\documentclass{aastex62}

\makeatletter
\newcommand*{\rom}[1]{\expandafter\@slowromancap\romannumeral #1@}
\makeatother

\received{\today}
\submitjournal{ApJL}

\shorttitle{Radio Background}
\shortauthors{Dowell \& Taylor}

\begin{document}

\title{The Radio Background Below 100 MHz}

\correspondingauthor{Jayce Dowell}
\email{jdowell@unm.edu}

\author{Jayce Dowell}
\affil{University of New Mexico \\
1919 Lomas Blvd NE \\
Albuquerque, NM, 87131, USA}

\author{Greg B. Taylor}
\affil{University of New Mexico \\
1919 Lomas Blvd NE \\
Albuquerque, NM, 87131, USA}

\begin{abstract}
The recent detection of the ``cosmic dawn'' redshifted 21 cm signal at 78 MHz by the EDGES experiment differs significantly from theoretical predictions.  In particular, the absorption trough is roughly a factor of two stronger than the most optimistic theoretical models. The early interpretations of the origin of this discrepancy fall into two categories.  The first is that there is increased cooling of the gas due to interactions with dark matter, while the second is that the background radiation field includes a contribution from a component in addition to the cosmic microwave background.  In this paper we examine the feasibility of the second idea using new data from the first station of the Long Wavelength Array.  The data span 40 to 80 MHz and provide important constraints on the present-day background in a frequency range where there are few surveys with absolute temperature calibration suitable for measuring the strength of the radio monopole.  We find support for a strong, diffuse radio background that was suggested by the ARCARDE 2 results in the 3 to 10 GHz range.  We find that this background is well modeled by a power law with a spectral index of --2.58$\pm$0.05 and a temperature at the rest frame 21 cm frequency of 603$^{+102}_{-92}$ mK.
\end{abstract}


\section{Introduction}
The detection of the redshifted 21 cm signal from the period when the Universe transitioned from a predominately neutral medium to an ionized one has been the target of numerous experiments, such as \citet{PAPER}, \citet{LEDA}, and \citet{LOFAREOR} to name a few, over the last several years.  This period in the history of the Universe provides important constraints on the first stars and galaxies in the Universe \citep[see][for an overview of the signal]{madau97,fur06,21Cosmology}.  The claimed detection of this signal by \citet{bowman2018} challenges the existing theoretical predictions both by being significantly stronger than predicted (500 mK) and by having a non-Gaussian shape.  It has been suggested that the observed depth can be explained by additional cooling of the neutral hydrogen gas by interactions with dark matter \citep{burkana2018} which allows the gas to cool at a faster rate than with cosmological expansion alone.  This, in turn, allows the gas to be colder than expected by the time the first stars form.  The details of this idea have been further refined by \citet{munoz2018} and \citet{berlin2018}.

An alternative to additional cooling through interactions with dark matter is to enhance the background radiation field by including a component in addition to the cosmic microwave background \citep[CMB,][]{feng2018,mirocha2018}.  In particular, radiation from primordial back holes could provide a sufficient background field to produce the observed line depth without violating the constraints set by the diffuse cosmic backgrounds at other wavelengths \citep{ewall2018}.  The existence of such a radio background has been suggested for a number of years, e.g., \citet{bridle1967}, and gained traction with the ARCADE 2 results \citep{fixen2011,seiffert2011}.  In fact, the nature of this seemingly isotropic background was the focus of a recent workshop to discuss whether or not the background is real and what could explain it \citep[see][and references therein for an overview]{singal2018}.  However, the exact nature and interpretation of this background, also referred to as the radio synchrotron background, is sensitive to its spectrum, specifically, whether or not the background can be represented by a pure power law.

Low frequency observations, particularly below 100 MHz where the background appears to be roughly comparable to the Galactic foreground, provide the best constraints on the strength and spectrum of the background.  In order to determine the strength of the radio background monopole, data with an absolute calibration of the zero level of the temperature scale are required.  To date, spectral analyses of the radio background, e.g., \citet{fixen2011}, have been able to rely on only a few existing surveys below 1 GHz that meet these criteria.  In addition to the limited sampling in this frequency range, the existing maps do not provide rigorous analysis of the uncertainties in the maps, either in the two dimensional structure nor in the accuracy of the absolute calibration.  This makes it difficult to disentangle the radio background from any calibration errors.

To address these limitations in existing low frequency maps we have studied the radio background between 40 and 80 MHz using the all-sky maps from the LWA1 Low Frequency Sky Survey \citep[LLFSS,][]{skysurvey}.  These maps have the advantage of a consistent calibration of the absolute zero level across the survey and probe the background in a frequency range where few other absolutely calibrated maps are available.  In \S\ref{sec:data} we present our analysis method for measuring the background for a combination of LLFSS and literature data.  Section \ref{sec:results} presents the results on the spectrum of the radio background and our conclusions are presented in \S\ref{sec:conclusions}.

\section{Data and Methods}
\label{sec:data}

\subsection{Data}
We used a combination of literature data and data from the LLFSS in our analysis of the radio background.  From the literature we used three low frequency, $<$ 1 GHz, radio surveys:  the 22 MHz survey of \citet{GSM22}, the 45 MHz survey of \citet{GSM45A} and \citet{GSM45M}, and the re-processed 408 MHz survey of \citet{GSM408} and \citet{eHaslam}.  Above 1 GHz we have included data at 1.4 GHz from \citet{GSM14201}, \citet{GSM14202}, and \citet{GSM14203} and the ARCADE 2 data at six frequencies spanning 3.1 to 10 GHz \citep{singal2011}.  With the exception of the ARCADE 2 data, all of these maps cover $>$50\% of the sky and the data are provided in antenna temperature.

From the LLFSS we have included the 40, 50, 60, 70, and 80 MHz maps.  These maps include the radio sky down to a declination of $-$40$^\circ$ at a spatial resolution of a few degrees (4.1$^\circ$ at 40 MHz, 2.0$^\circ$ at 80 MHz) and pixel-to-pixel estimates of the uncertainties.  The absolute calibration for the LLFSS is provided by special dipoles that were developed for the Large aperture Experiment to detect the Dark Ages \citep[LEDA;][]{LEDA}.  These dipoles have an integrated three-state temperature calibration system that allows absolute sky temperatures to be found to better than 10 K over 40 to 80 MHz (see \citealt{FL} and \citealt{skysurvey} for details).  For the analysis we have chosen not to include the 35, 38, 45, or 74 MHz maps from the LLFSS since they do not provide significant additional constraints to the spectrum of the radio background.

\subsection{Galactic Modeling}
\label{sec:modeling}
In order to estimate the temperature of the background, the foreground Galactic emission must be removed.  Following the methods of \citet{kogut2011}, we have used two different approaches to analyze the foreground along three lines of sight:  the north and south Galactic poles and the coldest area in the northern Galactic hemisphere, the region around $b=+48^\circ$, $l=196^\circ$.  We will describe both methods briefly here.  In the first technique a simple plane-parallel model is used to estimate the foreground temperature along the lines of sight to the Galactic polar caps, i.e., $|b| > 75^\circ$.  For the plane-parallel model we averaged over Galactic longitude to suppress various Galactic structures, such as the North Polar Spur and other radio loops, and fit latitudes greater than 10$^\circ$ with a simple cosecant model of

\begin{equation}
T(\nu) = T_0(\nu) + T_G(\nu) \times \csc{}|b|\mbox{.}
\end{equation}

\noindent Here $T$ is the observed temperature, $T_G$ is the temperature of the Galactic foreground at the pole, and $T_0$ is a constant offset that includes the radio background.  We independently fit the northern and southern Galactic hemisphere to estimate the temperature of the Galactic contribution at the respective poles.  We do not estimate the foreground in the direction of the northern Galactic cold spot since its location is not consistent with a simple plane parallel model, i.e., the coldest areas should be located at the poles in this model.

\begin{figure}
	\centering
    \includegraphics{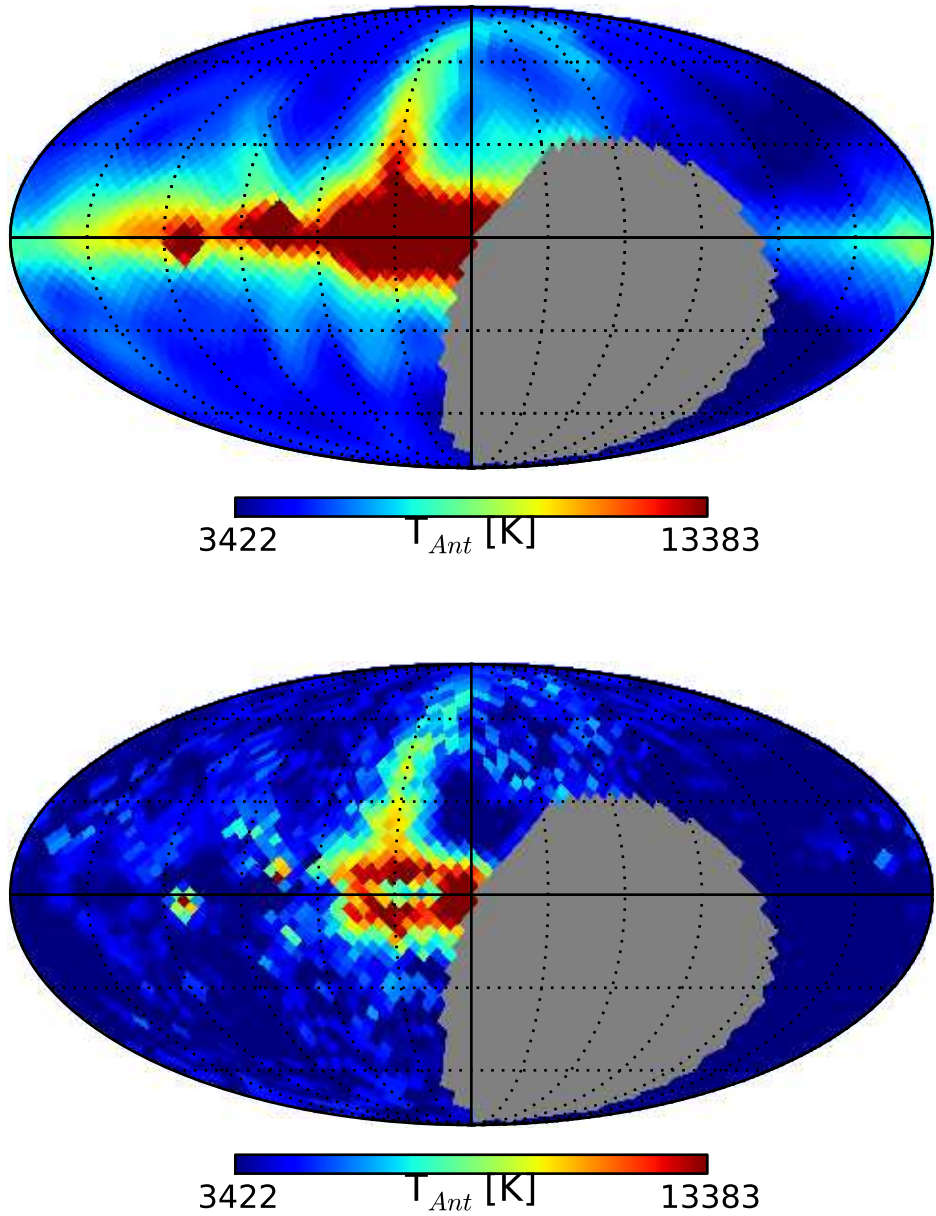}
    \caption{\label{fig:radiocorr}Example of the radio/C\rom{2} correlation applied to the 50 MHz LLFSS map.  The top panel shows the initial LLFSS map smoothed to a resolution of 7$^\circ$ to match the COBE/FIRAS data while the bottom panel shows the results of removing the Galactic foreground using the correlation with the 158$\mu$m C\rom{2} line.  This process removed most of the Galaxy except in the directions of the North Polar Spur and the Galactic center.}
\end{figure}

The second method of modeling the Galactic emission uses the correlation between the strengths of the synchrotron emission and the collisionally excited C\rom{2} line to remove the two dimensional foreground structure.  This is done by using the COBE/FIRAS maps of the C\rom{2} line at 158$\mu$m \citep{bennett1994} and fitting a function of the form

\begin{equation}
T(\nu) = T_0(\nu) + a(\nu) \times \sqrt{I_{CII}}\mbox{,}
\end{equation}

\noindent where $I_{CII}$ is the strength of the C\rom{2} line.  The radio and C\rom{2} data fall into two families and we fit both using an iterative method that assigns points to each family as was done by \citet{kogut2011}.  We then average the best fit relationships from each family to arrive at the final radio/C\rom{2} correlation.  The Galactic contribution to the temperature in a region is then

\begin{equation}
\langle T_G(\nu)\rangle = a(\nu) \langle \sqrt{I_{CII}}\rangle\mbox{,}
\end{equation}

\noindent where $\langle\rangle$ denotes a spatial average over the region.  This method allows us to not only estimate the Galactic temperature at the poles but also along the third line of sight in the direction of the coldest area of the Galactic northern hemisphere.  Figure \ref{fig:radiocorr} shows an example of the radio/C\rom{2} correlation technique applied to the 50 MHz LLFSS map.  Outside of the inner Galaxy and the North Polar Spur the map is relatively flat and featureless.

\begin{figure}
	\centering
    \includegraphics{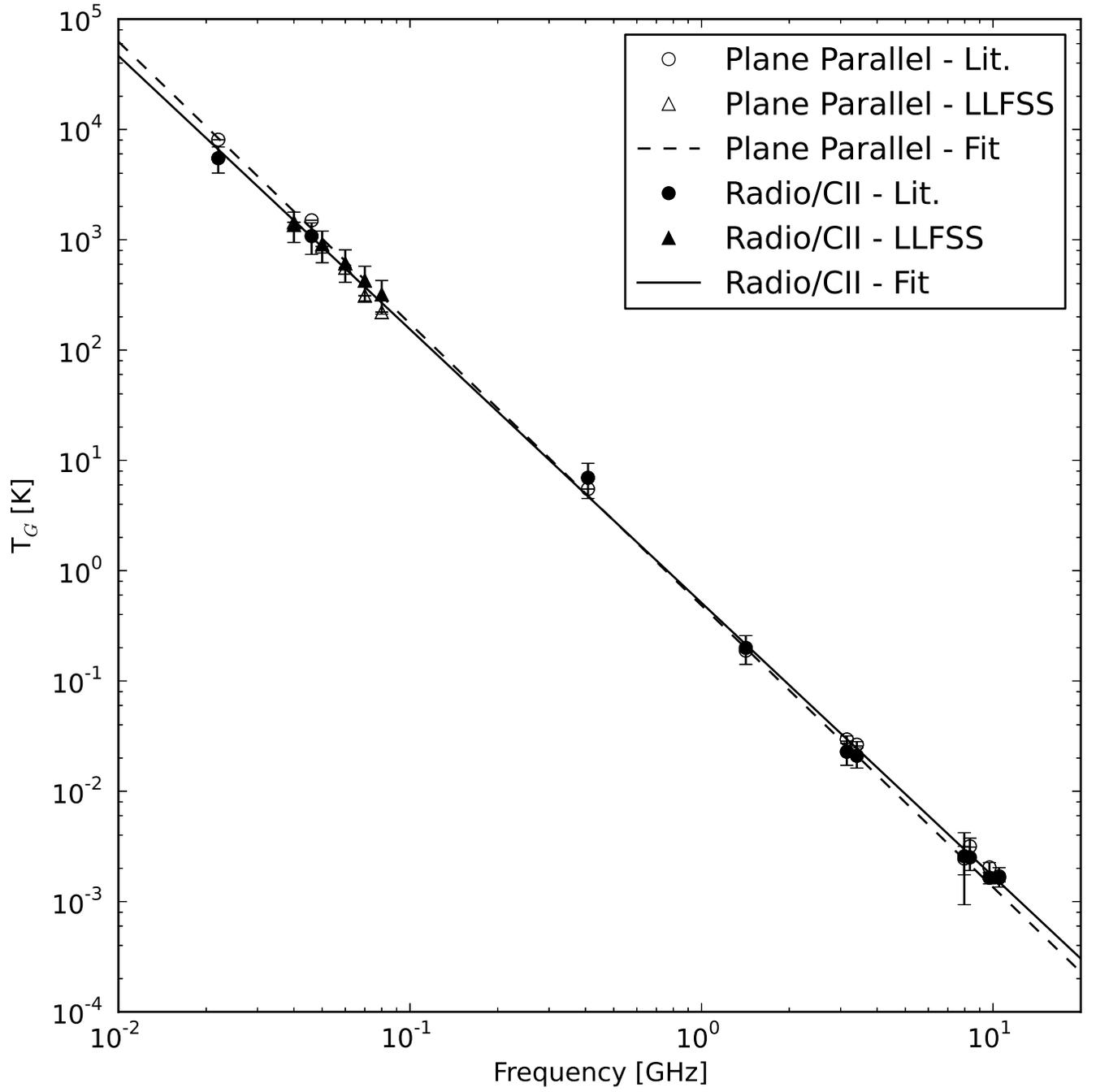}
    \caption{\label{fig:north}Models for the temperature of the Galactic emission in the direction of the north Galactic pole, $b>75^\circ$.  The triangles are obtained from LLFSS maps while the circles are those obtained from the other maps listed in Table \ref{tab:galactic}.  The open symbols are derived from a simple plane-parallel model while the filled circles come from a radio-C\rom{2} correlation.  Both models are consistent with a pure power law for the Galactic emission with a spectral index of $\approx$--2.55.}
\end{figure}

The results of these modeling procedures are fit with a power law of 

\begin{equation}
T_G(\nu) = T_{Gal} \left(\nu/\nu_0\right)^{\beta_{Gal}}\mbox{,}
\end{equation}

\noindent where $T_{Gal}$ is the strength of the emission at a reference frequency $\nu_0$ and $\beta_{Gal}$ is the spectral index.  The best fit power laws are presented in Table \ref{tab:galactic}, and Figure \ref{fig:north} shows the results for the north Galactic polar cap.  In general we find that our spectral indices are slightly shallower than those of \citet{kogut2011}, but are consistent within the errors.  We also find that the uncertainties of our parameters are slightly smaller which is likely due to differences in the assumed uncertainties in the underlying map data.  We also see good agreement between the fits when using all of the available data and when we exclude the 22 and 45 MHz data.

\section{Results}
\label{sec:results}
By removing the Galactic synchrotron using the models found in \S\ref{sec:modeling} we can estimate the level of the radio background.  The three lines of sight and two methods used result in a total of five measurements of the background at each frequency.  We combine these sets of five values using a simple average and estimate the uncertainty using the standard deviation of the measurements.  Table \ref{tab:background} lists the resulting thermodynamic radio background temperature between 22 MHz and 10.49 GHz.  These values are in good agreement with Table 4 of \citet{fixen2011}.  We fit these data using a combination of the CMB temperature and a power law of the form

\begin{equation}
T(\nu) = T_{CMB} + T_{R}\left(\nu/\nu_0\right)^\beta_{R}\mbox{,}
\end{equation}

\noindent where $\nu_0$ is 310 MHz.  We find a CMB temperature of 2.722$\pm$0.022 K, a spectral index of --2.58$\pm$0.05, and a background temperature, $T_R$, of 30.4$\pm$2.6 K at 310 MHz.

The most obvious difference between our results and the literature is in the temperature of the radio background at 310 MHz which we find to be $\approx$6 K higher than the best-fit value of \citet{fixen2011} of 24.1$\pm$2.1 K.  This is likely a result of systematic offsets in the zero points of the various low frequency surveys.  Indeed, if we exclude the LLFSS points and re-fit the data we find background temperature of 25.8$\pm$1.8 K and a spectral index of --2.56$\pm$0.04, consistent with \citet{fixen2011}.  To avoid the problems associated with varying zero levels between different surveys we re-fit using only the LWA1 data below 100 MHz since these data share a common instrument and calibration.  The result of this fit is similar to the results from the full data set.  Specifically, the temperature of the CMB is 2.732$\pm$0.013, the spectral index steepens to --2.66$\pm$0.04, and background temperature becomes 28.5$\pm$1.7 K at 310 MHz.

These power law components correspond to 0.603$^{+0.102}_{-0.092}$ K for the full data set and 0.497$^{+0.062}_{-0.056}$ K for the reduced low frequency data set at the rest frame frequency of the 21 cm transition.  The value for the radio background temperature obtained from the full fit is approximately 100 mK larger and with a marginally shallower spectral index than was obtained by \citet{fixen2011} while the LWA1 fit is within 20 mK albeit with a steeper spectral index.  For the remainder of the discussion we focus on the results from the full data set.  Of this background, approximately 100 mK can be explained by known source populations \citep[see][for a discussion of the known source classes]{singal2018}.  Thus, this leaves $\approx$500 mK unaccounted for.  If some of this background is produced at high redshifts then it could result in a significant enhancement of the background radiation field in addition to the CMB. \citet{feng2018} note that only a small fraction of this excess, on the order of a few percent, is needed in the redshift range of $z=$15 to 20 in order to produce the enhanced depth of the redshifted 21 cm line.  However, \citet{mirocha2018} caution that an enhanced radiation field has implications for the star formation and galaxy evolution processes at high redshifts that need to be reconciled in order to fully understand the underlying physics.

We note that these results are sensitive to the exact details of the removal of the Galactic contribution to the emission.  For example, \citet{fornengo2014} use a model of the Galactic magnetic field and the relativistic electron population to estimate the synchrotron emission.  Their model predicts more synchrotron toward the lower frequencies and reduces the radio background level by about a factor of two at 22 MHz.  Similarly, if we adjust the Galactic fitting procedure in \S\ref{sec:modeling} by excluding the plane-parallel model, not averaging the two families in the radio/C\rom{2} correlation, and excluding the northern polar cap, we find a shallower spectral index than before.  The best fit to the data in this case has a background temperature of 26.6$\pm$2.5 K at 310 MHz and a spectral index of --2.44$\pm$0.05.  This shallower slope is consistent with the results of \citet{fornengo2014} and \citet{vasilenko2017} but does not dramatically alter the temperature at 1.4 GHz of 0.654$^{+0.113}_{-0.102}$ K.  Furthermore, we note that this still requires a separate source of radiation in addition to the expected background from known source counts.

\section{Conclusions}
\label{sec:conclusions}
We have used data from the recently completed LWA1 Low Frequency Sky Survey to examine the existence of an isotropic radio background that cannot be explained by known populations of sources.  The origin of this background is of particular interest given the recent claim of a detection of the ``cosmic dawn'' signal which shows an unexpectedly strong absorption feature.  A potential explanation for the excess absorption is that the background radiation field includes a component in addition to the CMB which boosts the depth of the line.  Using our data we find results consistent with the previous analysis of the ARCADE 2 data and other zero level calibrated low frequency map data, lending support to the existence of a radio background.  This suggests that if the background forms early in the history of the Universe it may be responsible for the observed 21 cm signal.  We also find that the background is consistent with a single power law and shows no indication of a break or turn over between 22 MHz and 3 GHz.

We have also found that the exact strength and spectral slope of the radio background is sensitive to how the Galactic foregrounds are modeled and removed.  However, we do not find that the observed background at 1.4 GHz can be lowered to a level where it can be accounted for by known source classes by changes to the modeling procedure.  Rather these changes to the modeling methods have the greatest impact on the value of the spectral index and introduce additional uncertainty into this value.  Additional data in the radio below 1 GHz would be helpful in understanding the Galactic foregrounds.  Similarly, physically motivated models of the sky, such as \citet{gmoss}, may also help provide a robust method of removing the foregrounds.  Finally, additional theoretical work on the potential sources of this background, such as supernovae of population \rom{3} stars \citep{bierman2014} or cluster mergers \citep{fang2016}, is needed in order to characterize the expected strengths, spectra, and formation times.

\acknowledgements{We thank the anonymous referee for a prompt and thorough reading of the manuscript.  Construction of the LWA has been supported by the Office of Naval Research under Contract N00014-07-C-0147. Support for operations and continuing development of the LWA1 was provided by the National Science Foundation under grant AST-1139974 of the University Radio Observatory program.}




\clearpage
\begin{deluxetable}{llcccccc}
	\tablecolumns{8}
    \tablecaption{Power Law Fits for the Galactic Foreground Models\label{tab:galactic}}
    \tablehead{
		\colhead{Parameter} & \colhead{Model} & \multicolumn{3}{c}{All Data} & \multicolumn{3}{c}{LWA1\tablenotemark{a}} \\
        \cline{3-5}
        \cline{6-8}
		\colhead{~} & \colhead{~} & \colhead{North} & \colhead{South} & \colhead{Coldest} & \colhead{North} & \colhead{South} & \colhead{Coldest} \\
}
\startdata
~                     & CII  & 9.32$\pm$0.47 & 8.62$\pm$0.49 & 8.87$\pm$0.50 & 9.66$\pm$0.62 & 9.13$\pm$0.63 & 9.20$\pm$0.66 \\
T$_{Gal}$(310 MHz) [K] & $\csc{}b$ & 9.65$\pm$0.39 & 8.06$\pm$0.50 & \nodata & 9.10$\pm$0.35 & 8.15$\pm$0.41 & \nodata \\
~                     & Mean\tablenotemark{b} & 9.52$\pm$0.30 & 8.34$\pm$0.35 & 8.87$\pm$0.50 & 9.24$\pm$0.31 & 8.45$\pm$0.35 & 9.20$\pm$0.66 \\ 
\hline
~       & CII & -2.48$\pm$0.02 & -2.50$\pm$0.02 & -2.52$\pm$0.02 & -2.49$\pm$0.02 & -2.52$\pm$0.02 & -2.53$\pm$0.02 \\
$\beta_{Gal}$ & $\csc{}b$ & -2.56$\pm$0.02 & -2.58$\pm$0.04 & \nodata & -2.51$\pm$0.03 & -2.62$\pm$0.04 & \nodata \\
~       & Mean\tablenotemark{b} & --2.51$\pm$0.01 & --2.52$\pm$0.02 & --2.52$\pm$0.02 & --2.50$\pm$0.02 & --2.54$\pm$0.02 & --2.53$\pm$0.02 \\ 
\enddata
\tablenotetext{a}{Fit excluding the 22 and 45 MHz data.}
\tablenotetext{b}{Variance-weighted mean of the foreground estimation methods.}
\end{deluxetable}

\clearpage
\begin{deluxetable}{llcc}
	\tablecolumns{4}
    \tablecaption{Mean Extragalactic Background Temperature\label{tab:background}}
    \tablehead{
    	\colhead{Frequency} & \colhead{Map Source} & \colhead{Temperature\tablenotemark{a}} & \colhead{Uncertainty\tablenotemark{b}} \\
        \colhead{[GHz]} & \colhead{~} & \colhead{[K]} & \colhead{[K]} \\
    }
\startdata
0.022  & \citet{GSM22}     & 19212 & 4095 \\
0.040  & LLFSS; \citet{skysurvey} & 5792  & 963 \\
0.046  & \citet{GSM45A,GSM45M} & 4090  & 691 \\
0.050  & LLFSS; \citet{skysurvey} & 3443  & 526 \\
0.060  & LLFSS; \citet{skysurvey} & 2363  & 365 \\
0.070  & LLFSS; \citet{skysurvey} & 1505  & 208 \\
0.080  & LLFSS; \citet{skysurvey} & 1188  & 112 \\
0.408  & \citet{GSM408,eHaslam} & 15.20 & 2.37 \\
1.419  & \citet{GSM14201,GSM14202,GSM14203} & 3.276 & 0.167 \\
3.150  & ARCADE 2; \citet{fixen2011} & 2.788 & 0.045 \\
3.410  & ARCADE 2; \citet{fixen2011} & 2.768 & 0.045 \\
7.970  & ARCADE 2; \citet{fixen2011} & 2.764 & 0.060 \\
8.330  & ARCADE 2; \citet{fixen2011} & 2.741 & 0.062 \\
9.720  & ARCADE 2; \citet{fixen2011} & 2.731 & 0.062 \\
10.490 & ARCADE 2; \citet{fixen2011} & 2.731 & 0.065 \\
\enddata
\tablenotetext{a}{The temperatures shown here have been converted to from antenna to thermodynamic temperature.}
\tablenotemark{b}{Estimated from the five different combinations of Galaxy removal method and line-of-sight.}
\end{deluxetable}

\clearpage

\clearpage

\clearpage
\begin{figure}
	\centering
    \includegraphics{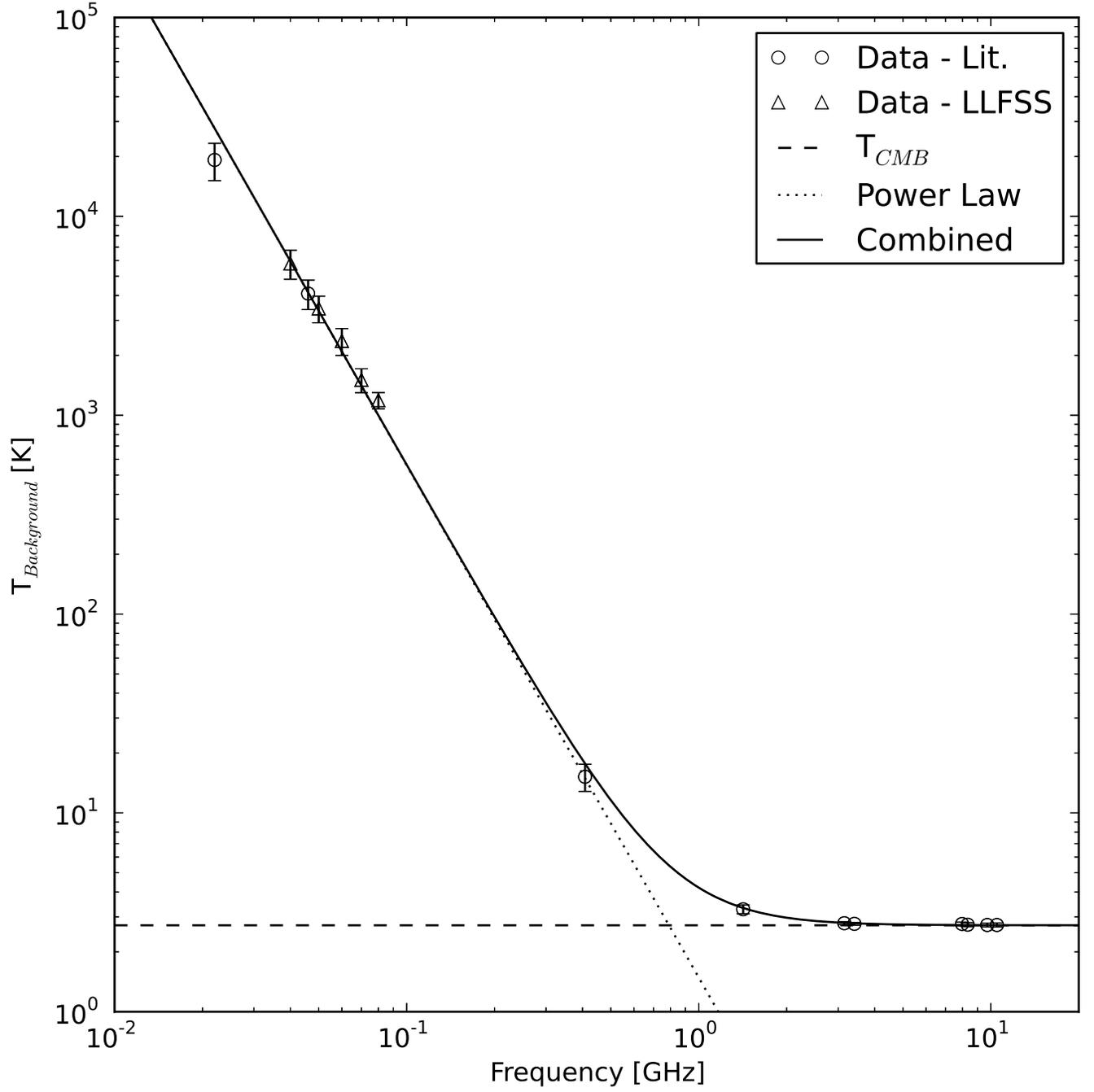}
    \caption{\label{fig:background}Modeled extragalactic temperature as a function of frequency.  The triangles are obtained from LLFSS maps while the circles are those obtained from the other maps listed in Table \ref{tab:galactic}.  The solid line shows the best fit to the sum of a power law (dotted line) with a spectral index of --2.58 and the CMB (dashed line) at 2.722 K.}
\end{figure}


\begin{thebibliography}{}
\expandafter\ifx\csname natexlab\endcsname\relax\def\natexlab#1{#1}\fi
\providecommand{\url}[1]{\href{#1}{#1}}

\bibitem[{{Alvarez} {et~al.}(1997){Alvarez}, {Aparici}, {May}, \&
  {Olmos}}]{GSM45A}
{Alvarez}, H., {Aparici}, J., {May}, J., \& {Olmos}, F. 1997, \aaps, 124,
  doi:10.1051/aas:1997196

\bibitem[{{Barkana}(2018)}]{burkana2018}
{Barkana}, R. 2018, \nat, 555, 71

\bibitem[{{Bennett} {et~al.}(1994){Bennett}, {Fixsen}, {Hinshaw}, {Mather},
  {Moseley}, {Wright}, {Eplee}, {Gales}, {Hewagama}, {Isaacman}, {Shafer}, \&
  {Turpie}}]{bennett1994}
{Bennett}, C.~L., {Fixsen}, D.~J., {Hinshaw}, G., {et~al.} 1994, \apj, 434, 587

\bibitem[{{Berlin} {et~al.}(2018){Berlin}, {Hooper}, {Krnjaic}, \&
  {McDermott}}]{berlin2018}
{Berlin}, A., {Hooper}, D., {Krnjaic}, G., \& {McDermott}, S.~D. 2018, ArXiv
  e-prints, arXiv:1803.02804

\bibitem[{{Biermann} {et~al.}(2014){Biermann}, {Nath}, {Caramete}, {Harms},
  {Stanev}, \& {Becker Tjus}}]{bierman2014}
{Biermann}, P.~L., {Nath}, B.~B., {Caramete}, L.~I., {et~al.} 2014, \mnras,
  441, 1147

\bibitem[{{Bowman} {et~al.}(2018){Bowman}, {Rogers}, {Monsalve}, {Mozdzen}, \&
  {Mahesh}}]{bowman2018}
{Bowman}, J.~D., {Rogers}, A.~E.~E., {Monsalve}, R.~A., {Mozdzen}, T.~J., \&
  {Mahesh}, N. 2018, \nat, 555, 67

\bibitem[{{Bridle}(1967)}]{bridle1967}
{Bridle}, A.~H. 1967, \mnras, 136, 219

\bibitem[{{Dowell} {et~al.}(2017){Dowell}, {Taylor}, {Schinzel}, {Kassim}, \&
  {Stovall}}]{skysurvey}
{Dowell}, J., {Taylor}, G.~B., {Schinzel}, F.~K., {Kassim}, N.~E., \&
  {Stovall}, K. 2017, \mnras, 469, 4537

\bibitem[{{Ewall-Wice} {et~al.}(2018){Ewall-Wice}, {Chang}, {Lazio},
  {Dor{\'e}}, {Seiffert}, \& {Monsalve}}]{ewall2018}
{Ewall-Wice}, A., {Chang}, T.-C., {Lazio}, J., {et~al.} 2018, ArXiv e-prints,
  arXiv:1803.01815

\bibitem[{Fang \& Linden(2016)}]{fang2016}
Fang, K., \& Linden, T. 2016, Journal of Cosmology and Astroparticle Physics,
  2016, 004.
\newblock \url{http://stacks.iop.org/1475-7516/2016/i=10/a=004}

\bibitem[{{Feng} \& {Holder}(2018)}]{feng2018}
{Feng}, C., \& {Holder}, G. 2018, ArXiv e-prints, arXiv:1802.07432

\bibitem[{{Fixsen} {et~al.}(2011){Fixsen}, {Kogut}, {Levin}, {Limon}, {Lubin},
  {Mirel}, {Seiffert}, {Singal}, {Wollack}, {Villela}, \&
  {Wuensche}}]{fixen2011}
{Fixsen}, D.~J., {Kogut}, A., {Levin}, S., {et~al.} 2011, \apj, 734, 5

\bibitem[{Fornengo {et~al.}(2014)Fornengo, Lineros, Regis, \&
  Taoso}]{fornengo2014}
Fornengo, N., Lineros, R.~A., Regis, M., \& Taoso, M. 2014, Journal of
  Cosmology and Astroparticle Physics, 2014, 008.
\newblock \url{http://stacks.iop.org/1475-7516/2014/i=04/a=008}

\bibitem[{{Furlanetto} {et~al.}(2006){Furlanetto}, {Oh}, \& {Briggs}}]{fur06}
{Furlanetto}, S.~R., {Oh}, S.~P., \& {Briggs}, F.~H. 2006, \physrep, 433, 181

\bibitem[{{Greenhill} \& {Bernardi}(2012)}]{LEDA}
{Greenhill}, L.~J., \& {Bernardi}, G. 2012, ArXiv e-prints, arXiv:1201.1700

\bibitem[{{Haslam} {et~al.}(1982){Haslam}, {Salter}, {Stoffel}, \&
  {Wilson}}]{GSM408}
{Haslam}, C.~G.~T., {Salter}, C.~J., {Stoffel}, H., \& {Wilson}, W.~E. 1982,
  \aaps, 47, 1

\bibitem[{{Kogut} {et~al.}(2011){Kogut}, {Fixsen}, {Levin}, {Limon}, {Lubin},
  {Mirel}, {Seiffert}, {Singal}, {Villela}, {Wollack}, \&
  {Wuensche}}]{kogut2011}
{Kogut}, A., {Fixsen}, D.~J., {Levin}, S.~M., {et~al.} 2011, \apj, 734, 4

\bibitem[{{Madau} {et~al.}(1997){Madau}, {Meiksin}, \& {Rees}}]{madau97}
{Madau}, P., {Meiksin}, A., \& {Rees}, M.~J. 1997, \apj, 475, 429

\bibitem[{{Maeda} {et~al.}(1999){Maeda}, {Alvarez}, {Aparici}, {May}, \&
  {Reich}}]{GSM45M}
{Maeda}, K., {Alvarez}, H., {Aparici}, J., {May}, J., \& {Reich}, P. 1999,
  \aaps, 140, 145

\bibitem[{Mirocha \& Furlanetto(2018)}]{mirocha2018}
Mirocha, J., \& Furlanetto, S.~R. 2018, \mnras, \textit{submitted}, arXiv:1803.03272

\bibitem[{{Mu{\~n}oz} \& {Loeb}(2018)}]{munoz2018}
{Mu{\~n}oz}, J.~B., \& {Loeb}, A. 2018, ArXiv e-prints, arXiv:1802.10094

\bibitem[{{Parsons} {et~al.}(2010){Parsons}, {Backer}, {Foster}, {Wright},
  {Bradley}, {Gugliucci}, {Parashare}, {Benoit}, {Aguirre}, {Jacobs},
  {Carilli}, {Herne}, {Lynch}, {Manley}, \& {Werthimer}}]{PAPER}
{Parsons}, A.~R., {Backer}, D.~C., {Foster}, G.~S., {et~al.} 2010, \aj, 139,
  1468

\bibitem[{{Pritchard} \& {Loeb}(2012)}]{21Cosmology}
{Pritchard}, J.~R., \& {Loeb}, A. 2012, Reports on Progress in Physics, 75,
  086901

\bibitem[{{Reich} \& {Reich}(1986)}]{GSM14202}
{Reich}, P., \& {Reich}, W. 1986, \aaps, 63, 205

\bibitem[{{Reich} {et~al.}(2001){Reich}, {Testori}, \& {Reich}}]{GSM14203}
{Reich}, P., {Testori}, J.~C., \& {Reich}, W. 2001, \aap, 376, 861

\bibitem[{{Reich}(1982)}]{GSM14201}
{Reich}, W. 1982, \aaps, 48, 219

\bibitem[{Remazeilles {et~al.}(2015)Remazeilles, Dickinson, Banday, Bigot-Sazy,
  \& Ghosh}]{eHaslam}
Remazeilles, M., Dickinson, C., Banday, A.~J., Bigot-Sazy, M.-A., \& Ghosh, T.
  2015, Monthly Notices of the Royal Astronomical Society, 451, 4311.
\newblock \url{+ http://dx.doi.org/10.1093/mnras/stv1274}

\bibitem[{{Roger} {et~al.}(1999){Roger}, {Costain}, {Landecker}, \&
  {Swerdlyk}}]{GSM22}
{Roger}, R.~S., {Costain}, C.~H., {Landecker}, T.~L., \& {Swerdlyk}, C.~M.
  1999, \aaps, 137, 7

\bibitem[{{Sathyanarayana Rao} {et~al.}(2017){Sathyanarayana Rao},
  {Subrahmanyan}, {Udaya Shankar}, \& {Chluba}}]{gmoss}
{Sathyanarayana Rao}, M., {Subrahmanyan}, R., {Udaya Shankar}, N., \& {Chluba},
  J. 2017, \aj, 153, 26

\bibitem[{{Seiffert} {et~al.}(2011){Seiffert}, {Fixsen}, {Kogut}, {Levin},
  {Limon}, {Lubin}, {Mirel}, {Singal}, {Villela}, {Wollack}, \&
  {Wuensche}}]{seiffert2011}
{Seiffert}, M., {Fixsen}, D.~J., {Kogut}, A., {et~al.} 2011, \apj, 734, 6

\bibitem[{Singal {et~al.}(2011)Singal, Fixsen, Kogut, Levin, Limon, Lubin,
  Mirel, Seiffert, Villela, Wollack, \& Wuensche}]{singal2011}
Singal, J., Fixsen, D.~J., Kogut, A., {et~al.} 2011, The Astrophysical Journal,
  730, 138.
\newblock \url{http://stacks.iop.org/0004-637X/730/i=2/a=138}

\bibitem[{{Singal} {et~al.}(2018){Singal}, {Haider}, {Ajello}, {Ballantyne},
  {Bunn}, {Condon}, {Dowell}, {Fixsen}, {Fornengo}, {Harms}, {Holder}, {Jones},
  {Kellermann}, {Kogut}, {Linden}, {Monsalve}, {Mertsch}, {Murphy}, {Orlando},
  {Regis}, {Scott}, {Vernstrom}, \& {Xu}}]{singal2018}
{Singal}, J., {Haider}, J., {Ajello}, M., {et~al.} 2018, \pasp, 130, 036001

\bibitem[{{Taylor} {et~al.}(2012){Taylor}, {Ellingson}, {Kassim}, {Craig},
  {Dowell}, {Wolfe}, {Hartman}, {Bernardi}, {Clarke}, {Cohen}, {Dalal},
  {Erickson}, {Hicks}, {Greenhill}, {Jacoby}, {Lane}, {Lazio}, {Mitchell},
  {Navarro}, {Ord}, {Pihlstr{\"o}m}, {Polisensky}, {Ray}, {Rickard},
  {Schinzel}, {Schmitt}, {Sigman}, {Soriano}, {Stewart}, {Stovall}, {Tremblay},
  {Wang}, {Weiler}, {White}, \& {Wood}}]{FL}
{Taylor}, G.~B., {Ellingson}, S.~W., {Kassim}, N.~E., {et~al.} 2012, Journal of
  Astronomical Instrumentation, 1, 1250004

\bibitem[{Vasilenko \& Sidorchuk(2017)}]{vasilenko2017}
Vasilenko, N.~M., \& Sidorchuk, M.~A. 2017, Astrophysics and Space Science,
  362, 221

\bibitem[{{Yatawatta} {et~al.}(2013){Yatawatta}, {de Bruyn}, {Brentjens},
  {Labropoulos}, {Pandey}, {Kazemi}, {Zaroubi}, {Koopmans}, {Offringa},
  {Jeli{\'c}}, {Martinez Rubi}, {Veligatla}, {Wijnholds}, {Brouw}, {Bernardi},
  {Ciardi}, {Daiboo}, {Harker}, {Mellema}, {Schaye}, {Thomas}, {Vedantham},
  {Chapman}, {Abdalla}, {Alexov}, {Anderson}, {Avruch}, {Batejat}, {Bell},
  {Bentum}, {Best}, {Bonafede}, {Bregman}, {Breitling}, {van de Brink},
  {Broderick}, {Br{\"u}ggen}, {Conway}, {de Gasperin}, {de Geus}, {Duscha},
  {Falcke}, {Fallows}, {Ferrari}, {Frieswijk}, {Garrett}, {Griessmeier},
  {Gunst}, {Hassall}, {Hessels}, {Hoeft}, {Iacobelli}, {Juette},
  {Karastergiou}, {Kondratiev}, {Kramer}, {Kuniyoshi}, {Kuper}, {van Leeuwen},
  {Maat}, {Mann}, {McKean}, {Mevius}, {Mol}, {Munk}, {Nijboer}, {Noordam},
  {Norden}, {Orru}, {Paas}, {Pandey-Pommier}, {Pizzo}, {Polatidis}, {Reich},
  {R{\"o}ttgering}, {Sluman}, {Smirnov}, {Stappers}, {Steinmetz}, {Tagger},
  {Tang}, {Tasse}, {ter Veen}, {Vermeulen}, {van Weeren}, {Wise}, {Wucknitz},
  \& {Zarka}}]{LOFAREOR}
{Yatawatta}, S., {de Bruyn}, A.~G., {Brentjens}, M.~A., {et~al.} 2013, \aap,
  550, A136

\end{thebibliography}
\end{document}